\documentclass[10pt,prd,twocolumn,showpacs,amsmath,amssymb,aps,floats,floatfix,nofootinbib]{revtex4-1}
\usepackage{amsmath,amsfonts,amssymb}
\usepackage{graphicx}
\usepackage{color}
\usepackage[dvipsnames,svgnames,x11names]{xcolor}
\usepackage{slashed} 
\usepackage{url}
\usepackage{subfigure}
\usepackage{multirow} 

\def\cm{\mathrm{cm}} 
\def\sec{\mathrm{s}} 
\def\ifb{\mathrm{fb}^{-1}} 
\def\GeV{\mathrm{GeV}} 
\def\pT{p_\mathrm{T}} 
\def\missET{\slashed E_\mathrm{T}} 
\def\missE{\slashed E} 
\def\tchi{{\tilde\chi}} 
\def\stau{{\tilde\tau}} 
\def\slep{{\tilde\ell}} 

\makeatother

\allowdisplaybreaks 

\begin{document}

\title{Tau Portal Dark Matter models at the LHC}
\author{Zhao-Huan Yu$^1$}
\author{Xiao-Jun Bi$^1$}
\author{Qi-Shu Yan$^{2,3}$}
\author{Peng-Fei Yin$^1$}
\affiliation{$^1$Key Laboratory of Particle Astrophysics,
Institute of High Energy Physics, Chinese Academy of Sciences,
Beijing 100049, China}
\affiliation{$^2$School of Physics,
University of Chinese Academy of Sciences,
Beijing 100049, China}
\affiliation{$^3$Center for High Energy Physics, Peking University, Beijing 100871, China}

\begin{abstract}
Motivated by the Galactic Center gamma-ray excess in the Fermi-LAT data, we study the signatures of a class of tau portal dark matter (DM) models where DM particles preferentially couple to tau leptons at the LHC. We consider the constraints from the DM direct detection and investigate the sensitivity of the LHC to di-tau plus missing energy signatures. We find that the LHC with a high luminosity of 3000 fb$^{-1}$ can test the tau portal DM models with fermionic mediators in the mass range of $120\sim450$ GeV.
\end{abstract}


\maketitle


\section{Introduction}

Various astrophysical and cosmological observations have confirmed that the main component of the matter in the Universe is non-baryonic dark matter. There is a class of attractive DM candidates called weakly interacting massive particles with masses of $\mathcal{O}(1)-\mathcal{O}(10^3)$~GeV. Their annihilations in regions with high DM densities can effectively generate energetic cosmic rays, such as gamma-rays, neutrinos, and antimatter particles, which could be captured by DM indirect detection experiments. Compared with charged particles, gamma-rays are less affected by the interstellar matter and the Galactic magnetic field during their propagation. Thus their energy spectra and directions as well as their origins can be well determined. Therefore, gamma-rays could be a very good probe to reveal the microscopic properties of DM particles in the Galaxy.

Recently, several groups reported an extended gamma-ray excess with a high significance peaking at a few GeV in the Fermi Large Area Telescope data~\cite{Goodenough:2009gk,Hooper:2010mq,Boyarsky:2010dr,Hooper:2011ti,Abazajian:2012pn,Macias:2013vya,Abazajian:2014fta,Daylan:2014rsa,Lacroix:2014eea,Zhou:2014lva,Calore:2014xka}. This excess emission is found in the Galactic Center (GC) and even a larger region up to $10^\circ$ from the GC after subtracting well-known astrophysical backgrounds. The spatial distribution of the excess is similar to the square of the Navarro-Frenk-White DM distribution~\cite{Navarro:1996gj}. A study showed that such an excess is not contributed by emissions from the Fermi bubbles~\cite{Daylan:2014rsa}. However, it might arise from other astrophysical sources, such as a population of millisecond pulsars \cite{Abazajian:2010zy,Yuan:2014rca} or high energy cosmic rays from the explosion in the GC \cite{Carlson:2014cwa,Petrovic:2014uda}. Nevertheless, whether these astrophysical sources can simultaneously interpret the total flux, energy spectrum, and spatial distribution of the excess is still debatable~\cite{Yuan:2014rca,Hooper:2013nhl,Cholis:2014noa}.

An attractive and economic explanation for the gamma-ray excess is DM annihilations, if only the DM distribution is a steep Navarro-Frenk-White profile $\sim r^{-\gamma}$ with an inner slope $\gamma\sim1.2$. DM particles may directly annihilate into $b$ quarks, light quarks, charged leptons~\cite{Goodenough:2009gk,Hooper:2010mq,Hooper:2011ti,Daylan:2014rsa}, or some new light particles~\cite{Abdullah:2014lla,Martin:2014sxa,Berlin:2014pya}, which further produce photons from their cascade decays and/or final state radiations. For instance, the gamma-ray excess can be interpreted by a model with DM particles annihilating into $b\bar{b}$ with a mass of $30 - 40$~GeV and a cross section of $\sim 10^{-26}~\cm^{-1}~\sec^{-1}$~\cite{Abazajian:2014fta,Daylan:2014rsa}, which can naturally yield the correct DM relic density.

If DM particles dominantly couple to $b$ quarks at the tree level, they would generate nuclear recoil signatures via loop effects for DM direct detection experiments, which are very sensitive to $\mathcal{O}(10)$ GeV DM particles~\cite{Izaguirre:2014vva,Agrawal:2014una,Agrawal:2014una}. Future underground experiments, such as XENON and LUX, will test these scenarios. DM annihilations into $b$ quarks in the Galaxy will also produce extra antiprotons. Considering the PAMELA antiproton measurement with null excess~\cite{Adriani:2008zq}, some works have found stringent constraints on such DM models accounting for the gamma-ray excess~\cite{Bringmann:2014lpa,Cirelli:2014lwa}. Nonetheless, it is necessary to point out that these constraints are subject to unavoidable astrophysical uncertainties from cosmic-ray propagation and solar modulation models.

As well known, collider searches for dark matter are complimentary to the direct and indirect detection experiments. At high energy colliders, DM particle productions can yield events with a significant large missing transverse energy ($\missET$). In order to explain the gamma-ray excess, models with an appropriate mediator are favored. It is interesting to notice that the required DM particle mass and annihilation cross section indicate that the mediator connecting DM and SM particles should be around  $\mathcal{O}(10^2)$~GeV or so, which seems within the searching capability of the LHC. Thus it is hopeful to observe direct productions of the mediators at the LHC. For recent works on this topic, interested readers can refer the papers \cite{Boehm:2014hva,Izaguirre:2014vva,Agrawal:2014una,Berlin:2014tja,Yu:2014pra,Guo:2014gra}.

In this work, we focus on a class of simplified tau portal DM models where DM particles preferentially couples to tau leptons. The $\sim\mathcal{O}(1)$~TeV leptophilic DM particles have been extensively studied to interpret the anomalous high energy cosmic-ray positrons and electrons. Some recent works have investigated the collider phenomenology of general leptophilic DM models with masses of $\mathcal{O}(10^2)$ GeV \cite{Bai:2014osa,Chang:2014tea,Freitas:2014jla}. For a pure DM annihilation channel to $\tau^+\tau^-$, the gamma-ray excess can be explained by the DM particle with a mass of $\sim 9$ GeV and an annihilation cross section of $\sim 5\times 10^{-27}~\cm^{-1}~\sec^{-1}$~\cite{Abazajian:2014fta}.

Compared with the $\mathcal{O}(10)$~GeV DM candidates dominantly interacting with $b \bar{b}$ and/or light quarks, the $\mathcal{O}(1)$~GeV leptophilic DM candidates could avoid the stringent constraints from the direct detection due to the loop-suppressed couplings to quarks~\cite{Kopp:2009et,Bai:2014osa,Chang:2014tea,Kopp:2014tsa} and the experimental threshold. Meanwhile, when compared with the leptophilic DM candidates dominantly interacting with $e^+ e^-$ and/or $\mu^+ \mu^-$, the tau portal DM candidates would circumvent the constraints from the GeV cosmic-ray $e^\pm$ measurements due to the softer $e^\pm$ spectrum from tau decays, which is usually hidden by the sizable solar modulation effect~\cite{Hooper:2012gq,Bergstrom:2013jra}.

According to traditional wisdom, it may pose a challenge to detect the tau portal DM particles at colliders for their vanishing couplings to the beam particles. Instead, we investigate the mediator productions in the tau portal DM models. The expected DM annihilation cross section to account for the gamma-ray excess require that the mediators connecting DM particles to taus are moderate heavy, saying several hundred GeV or lighter. These mediators are tau partners carrying the same electric and hyper charges and can be directly pair-produced via the Drell-Yan process. The final states of the signals could be $\tau^+ \tau^-$ plus a large missing transverse energy $\missET$. Therefore it is well motivated to study the sensitivity to the signatures of these models at the LHC.

This paper is organized as follows. In Sec.~\ref{sec:model}, we briefly introduce the simplified models containing DM particles and mediators. In Sec.~\ref{sec:DD}, we study the DM-nucleon scattering and derive the limits from the direct detection. In Sec.~\ref{sec:LHC}, we investigate the phenomenology of these tau portal models at the LHC. The sensitivity to these models is obtained with $\sqrt{s}=14$~TeV and integrated luminosities of $300~\ifb$ and $3000~\ifb$. Sec.~\ref{sec:concl} gives our conclusions and discussions.

\section{Tau portal Dark Matter models}
\label{sec:model}

In this section, we briefly describe a class of simplified models
in which only the tau lepton communicates with the dark sector.
The dark sector contains two particles, which are both required
to be odd in a $Z_2$ symmetry. Under such a symmetry, the tau lepton is assigned to be even. Consequently, the lighter particle in the dark sector, denoted as $\chi$, will be stable and become a DM candidate.
The heavier particle is a mediator connecting
the DM particle to the tau lepton.

Here we consider a few models in two cases: the case that the DM particle is a fermion with a spin 1/2 and the case that it is a scalar with a spin 0. In all these models, a pair of DM particles can annihilate into $\tau^+ \tau^-$ via the $t$-channel exchange of the mediator. For simplicity, we neglect those models involving DM particles with higher spins.

For the fermionic DM particle case, we consider two models, the Dirac and Majorana fermionic dark matter (DFDM and MFDM), respectively. In these models, the mediator is assumed to be a scalar $\phi$ and the dark sector
interacts with the right-handed tau via a renormalizable Yukawa-type interaction
\begin{equation}
\mathcal{L}_\phi =
\lambda \,\phi \bar\tau_R \chi_L + \mathrm{h.c.}\,.
\end{equation}
As $\chi$ is a singlet, $\phi$ should carry a hyper charge $-1$ and
hence an electric charge $-1$, and a tau lepton number $+1$.
Thus $\phi$ interacts with photons and $Z$ bosons.
In supersymmetric models, this mediator $\phi$ corresponds to the right-handed stau.

For the scalar DM particle case, we consider other two models, namely the complex and real scalar dark matter (CSDM and RSDM), respectively. The mediator is assumed to be a fermion $\psi$
and the renormalizable interaction between the dark sector
and the right-handed tau is assumed to be
\begin{equation}
\mathcal{L}_\psi =
\kappa \, \chi \bar\tau_R \psi_L + \mathrm{h.c.}\,.
\end{equation}
Here the mediator $\psi$ carries a hyper charge $-1$, an electric charge $-1$, and a tau lepton number $+1$.

For the sake of simplicity, we will not assume the left-handed tau as a portal particle. Constructing a model with mediators in $SU(2)_L$ doublets will have to introduce more degrees of freedom and make the current study more complicated. Interested readers can see Ref.~\cite{Hagiwara:2013qya} for the phenomenology of the neutralino DM with a mass of $\sim 10$ GeV and light staus. In addition, we will not assume that there are other light mediators connecting the DM to the electron or the muon. If such particles exist in a low energy scale, they would lead to contributions to the electron or muon magnetic dipole moments and lepton flavor violation decays. Studies on this topic can be found in Refs.~\cite{Bai:2014osa,Chang:2014tea,Agrawal:2014ufa,Kopp:2014tsa}.

In the following study, we adopt the DM particle mass and the annihilation cross section to $\tau^+ \tau^-$ as
\begin{eqnarray}
m_\chi &=& 9.43~(_{-0.52}^{+0.63}~\mathrm{stat.})~(\pm 1.2~\mathrm{sys.})~\GeV,
\\
\left< \sigma_\mathrm{ann}v \right> &=&
(0.51 \pm 0.24)\times 10^{-26}~\cm^3~\sec^{-1},
\label{eq:sv:GC}
\end{eqnarray}
which can fit the GC gamma-ray excess~\cite{Abazajian:2014fta}. In the low-velocity limit, the thermally averaged annihilation cross sections for $\chi\chi\to\tau^+\tau^-$ in the tau portal models are given as follows:
\begin{eqnarray}
&&\text{DFDM:~~} \frac{1}{2}\left<\sigma_\mathrm{ann}v\right> =
\frac{\lambda^4 m_\chi^2 \beta_\tau}{64\pi(m_\phi^2 + m_\chi^2 - m_\tau^2)^2}
\nonumber\\
&&\qquad\quad \simeq 5 \times 10^{-27}~\cm^3~\sec^{-1}
\left(\frac{\lambda}{m_\phi /179~\GeV}\right)^4,
\label{eq:sv:DFDM}\\
&&\text{MFDM:~~} \left<\sigma_\mathrm{ann}v\right> =
\frac{\lambda^4 m_\tau^2 \beta_\tau}{32\pi(m_\phi^2 + m_\chi^2 - m_\tau^2)^2}
\nonumber\\
&&\qquad\quad \simeq 5 \times 10^{-27}~\cm^3~\sec^{-1}
\left(\frac{\lambda}{m_\phi /93~\GeV}\right)^4,
\label{eq:sv:MFDM}\\
&&\text{CSDM:~~} \frac{1}{2}\left<\sigma_\mathrm{ann}v\right> =
\frac{\kappa^4 m_\tau^2 \beta_\tau^3}{32\pi(m_\psi^2 + m_\chi^2 - m_\tau^2)^2}
\nonumber\\
&&\qquad\quad \simeq 5 \times 10^{-27}~\cm^3~\sec^{-1}
\left(\frac{\kappa}{m_\psi /93~\GeV}\right)^4,
\label{eq:sv:CSDM}\\
&&\text{RSDM:~~} \left<\sigma_\mathrm{ann}v\right> =
\frac{\kappa^4 m_\tau^2 \beta_\tau^3}{4\pi(m_\psi^2 + m_\chi^2 - m_\tau^2)^2}
\nonumber\\
&&\qquad\quad \simeq 5 \times 10^{-27}~\cm^3~\sec^{-1}
\left(\frac{\kappa}{m_\psi /156~\GeV}\right)^4,
\label{eq:sv:RSDM}
\end{eqnarray}
where $\beta_\tau\equiv\sqrt{1-m_\tau^2/m_\chi^2}$,
and the approximations are obtained under the condition
$m_\tau \ll m_\chi \ll m_\phi,m_\psi$.
For non-self-conjugate dark matter, the factors of $1/2$
in Eqs.~\eqref{eq:sv:DFDM} and \eqref{eq:sv:CSDM} are introduced
to compensate the difference from self-conjugate dark matter,
so that these quantities can be directly connected to Eq.~\eqref{eq:sv:GC}.
For the MFDM, CSDM, and RSDM, the $s$-wave annihilation cross sections
are proportional to $m_\tau^2$.
As we can see from the approximations, to achieve the fixed cross section under the assumption of perturbation, the mediators are required to be light, saying $\mathcal{O}(10^2)$ GeV or so. Obviously, such light mediators are hopefully accessible in collider searches. The parameter spaces in the models consistent with
the annihilation cross section~\eqref{eq:sv:GC} within $1\sigma$ uncertainty
are denoted by green bands in Fig.~\ref{fig:constr}.

\section{Direct detection of tau portal Dark Matter models}
\label{sec:DD}

In these models, although DM particles do not directly couple to
quarks and gluons, they can couple to nuclei
through loop-induced electromagnetic form factors.
This may lead to direct detection signals.
In the DFDM, MFDM, and CSDM models, DM particles scatter off nuclei
via exchanging a photon at 1-loop level, as shown in Fig.~\ref{fig:fd_Dphi_DD}.
But, for the MFDM, the leading interaction corresponds to
an anapole moment operator, which leads to a cross section inaccessible
in current direct detection experiments~\cite{Bai:2014osa}.
Additionally, the leading interaction between the RSDM
and nuclei arises from 2-loop diagrams via exchanging two photons,
and make it more difficult for direct searches~\cite{Chang:2014tea}.

Despite these difficulties, limits from recent spin-independent direct detection experiments, such as LUX~\cite{Akerib:2013tjd}, may put some constraints on the DFDM and CSDM models. Below we address these constraints.

\begin{figure}[!htbp]
\centering
\includegraphics[width=.48\textwidth]{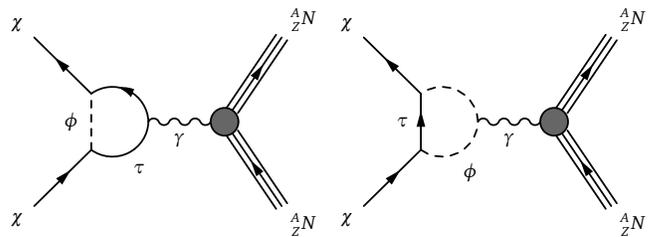}
\caption{1-loop contributions for the fermionic DM scattering off nucleus.
For the complex scalar DM, the diagrams are similar
after substituting the $\phi$ propagator with the $\psi$ propagator.}
\label{fig:fd_Dphi_DD}
\end{figure}

For the DFDM, the averaged DM-nucleon spin-independent scattering cross section
can be approximately expressed as~\cite{Agrawal:2011ze,Bai:2014osa}
\begin{equation}
\sigma_{\chi N} = \frac{Z^2 e^2 B^2 \mu_{\chi N}^2}{\pi A^2},
\end{equation}
where $\mu_{\chi N} \equiv m_\chi m_N/(m_\chi + m_N)$
is the DM-nucleon reduced mass,
and $A$ ($Z$) is the mass (atomic) number of the nucleus, and
\begin{equation}
B \simeq  - \frac{\lambda^2 e}{64\pi^2 m_\phi^2}
\left[\frac{1}{2} + \frac{2}{3}\ln\left(\frac{m_\tau^2}{m_\phi^2}\right)\right]
\end{equation}
is the form factor matched to the DM-photon effective operator
$[\bar\chi\gamma^\mu(1-\gamma_5)\partial^\nu\chi
+\mathrm{h.c.}]F_{\mu\nu}$.

For the CSDM, the averaged DM-nucleon spin-independent scattering cross section
is given by~\cite{Bai:2014osa}
\begin{equation}
\sigma_{\chi N} = \frac{Z^2 e^2 C^2\mu_{\chi N}^2}{8\pi A^2},
\end{equation}
where
\begin{equation}
C \simeq - \frac{\kappa^2 e}{16 \pi^2 m_\psi^2}
\left[ 1 + \frac{2}{3} \ln \left(\frac{m_\tau^2}{m_\psi^2}\right) \right]
\end{equation}
is the form factor matched to the DM-photon effective operator
$(\partial^\mu\chi)(\partial^\nu\chi^*)F_{\mu\nu}$.

For $m_\chi=9.43~\GeV$, the LUX experiment has excluded
$\sigma_{\chi N}\gtrsim 2\times 10^{44}~\cm^2$
at 90\% C.L.~\cite{Akerib:2013tjd}.
We use this bound to constrain the parameter spaces
of the DFDM and CSDM models, as shown
in Figs.~\ref{fig:constr:DFDM} and~\ref{fig:constr:CSDM}.
For the CSDM, the region consistent with the GC
gamma-ray excess has been already excluded.
For the DFDM, the LUX search has also excluded
a piece of the green band for $m_\phi \gtrsim 200~\GeV$,
and the remaining could be easily covered by
a near future search with larger exposure.

It should be noticed that current direct searches cannot constrain the MFDM and RSDM models.

\section{Collider searches}
\label{sec:LHC}

At $e^+e^-$ colliders, the $\phi\phi^*$ production signatures
for the DFDM and MFDM are essentially identical to
the signature of right-handed stau pair production in supersymmetric models.
Therefore, the bound on the right-handed stau from the LEP searches
up to $\sqrt{s}=208~\GeV$~\cite{Abdallah:2003xe}
can be applied to our cases.
This has excluded $m_\phi\lesssim 84~\GeV$.
As the $\psi\bar\psi$ production rate in $e^+e^-$ collisions
are higher than the $\phi\phi^*$ production rate by an order of magnitude,
we expect the LEP searches could exclude $m_\psi\lesssim 100~\GeV$
for the CSDM and RSDM. We display these bounds in Fig.~\ref{fig:constr}.

As the mediators $\phi$ and $\psi$ carry hyper and electric charges,
they can also be produced in pairs at the LHC via Drell-Yan processes,
i.e., exchanging an $s$-channel $\gamma/Z$.
Therefore, it is interesting to explore whether current and future LHC runs can directly discover these mediators or not.
The pair production cross sections for
$pp\to\phi\phi^*/\psi\bar\psi+\mathrm{jets}$
with collision energies of 8~TeV and 14~TeV are shown in Fig.~\ref{fig:Xsec}.
It is observed that for $m_\phi=m_\psi$, the cross sections of $\psi\bar\psi$ are roughly an order of magnitude larger than that of the $\phi\phi^*$ due to the reason that there are more helicity states for $\psi$ and the $\phi \phi^*$ production suffers a kinematic suppression in the angular distribution.

\begin{figure}[!htbp]
\centering
\includegraphics[width=.42\textwidth]{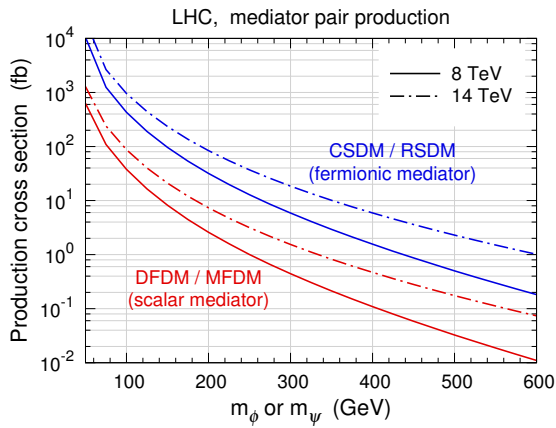}
\caption{Production cross sections for
$pp\to\phi\phi^*/\psi\bar\psi+\mathrm{jets}$
at the 8~TeV and 14~TeV LHC are shown.}
\label{fig:Xsec}
\end{figure}

Once produced, a mediator $\phi$ or $\psi$ is assumed to be 100\% decay into a tau and a DM particle. Taus can leptonically decay into $e\nu_e\nu_\tau$ and $\mu\nu_\mu\nu_\tau$ with branching ratios of 17.9\% and 17.4\%, respectively, and hadronically decay into the remaining modes with a branching ratio of 64.7\%.
The jets induced by hadronically decaying taus can be identified by
the $\tau$-tagging technique at the LHC.
Meanwhile, DM particles as well as neutrinos from tau decays will give rise to a missing transverse energy. Consequently, searches for the produced mediators can be classified into three channels: $2\tau_h+\missET$, $\tau_\ell\tau_h+\missET$, and $2\tau_\ell+\missET$,
where $\tau_h$ ($\tau_\ell$) denotes a hadronically (leptonically) decaying tau.

We generate background and signal simulation samples
using \texttt{MadGraph~5}~\cite{Alwall:2011uj},
to which the tau portal DM models is added
through \texttt{FeynRules~2}~\cite{Alloul:2013bka}.
\texttt{PYTHIA~6}~\cite{Sjostrand:2006za} is utilized to deal with
parton shower and hadronization processes,
as well as decay processes except for tau decays,
which is particularly performed by \texttt{TAUOLA}~\cite{Jadach:1993hs}.
The MLM matching scheme~\cite{Mangano:2001xp,Mangano:2006rw}
is adopted to avoid the double counting issue.
We use \texttt{Delphes~3}~\cite{deFavereau:2013fsa} to carry out
a parametrized fast detector simulation.
For the detector simulation at the 14~TeV LHC,
we adopt a set of parameters corresponding to the ATLAS detector.

At the LHC, the mediator productions are analogous to
some electroweak productions of supersymmetric particles,
for instance, $\tchi_1^+\tchi_1^-$, $\stau^+\stau^-$,
and $\slep^+\slep^-$ productions, which also have similar final states.
Such processes have been explored by recent experimental analyses,
e.g. Refs.~\cite{ATLAS:2013yla,Aad:2014yka,Aad:2014vma},
using the 8~TeV LHC data with an integrated luminosity of $\sim 20~\ifb$.
Based on our simulation, we have reinterpreted these experimental results,
and have found that they have yielded no meaningful constraints on the tau portal DM models.

Below, we discuss searches for the mediators at the 14~TeV LHC.
The $2\tau_h+\missET$ searching channel
significantly relies on the $\tau$-tagging technique.
According to the recent performance of the ATLAS detector
at 8~TeV~\cite{TheATLAScollaboration:2013wha},
two independent methods, boosted decision trees and projective likelihood,
give similar efficiencies for the discrimination of hadronically decaying taus.
For the medium (tight) $\tau$-tagging, the identification efficiency
is roughly 60\% (40\%), while the faking rate of QCD jets is about 5\% (2\%).
Although the medium $\tau$-tagging identifies more genuine taus,
its sizable faking rate would make it difficult to control
the enormous backgrounds with QCD jets. For this reason, we assume
an identification efficiency of 40\% and a faking rate of 2\%
in the simulation for the 14~TeV LHC.

For searching the signals of a pair of massive particles
which both decay into a visible object and an invisible object,
the ``stransverse'' mass $m_\mathrm{T2}$~\cite{Lester:1999tx,Barr:2003rg,Cheng:2008hk} has been found to be a useful variable
to discriminate signals from backgrounds. The observable $m_\mathrm{T2}$ is defined as
\begin{equation}
m_\mathrm{T2} = \min\limits_{\mathbf{p}_\mathrm{T}^1+
\mathbf{p}_\mathrm{T}^2=\slashed{\mathbf{p}}_\mathrm{T}}
\left\{\max\left[
m_\mathrm{T}(\mathbf{p}_\mathrm{T}^a,\mathbf{p}_\mathrm{T}^1),
m_\mathrm{T}(\mathbf{p}_\mathrm{T}^b,\mathbf{p}_\mathrm{T}^2)
\right]\right\},
\end{equation}
where $\mathbf{p}_\mathrm{T}^a$ and $\mathbf{p}_\mathrm{T}^b$
are the transverse momenta of the two visible objects,
which are reconstructed taus or light leptons in our cases.
$\mathbf{p}_\mathrm{T}^1$ and $\mathbf{p}_\mathrm{T}^2$ correspond to
a partition of the missing transverse momentum
$\slashed{\mathbf{p}}_\mathrm{T}$.
The transverse mass $m_\mathrm{T}$ is defined by
$m_\mathrm{T}(\mathbf{p}_\mathrm{T},\mathbf{q}_\mathrm{T})
=\sqrt{2(p_\mathrm{T}q_\mathrm{T}
-\mathbf{p}_\mathrm{T}\cdot\mathbf{q}_\mathrm{T})}$.
$m_\mathrm{T2}$ is the minimization of the larger $m_\mathrm{T}$
over all possible partitions.
The $m_\mathrm{T2}$ distribution has an upper endpoint
related to the masses of the semi-visible decaying particle
and the invisible particle.
For a large mass difference between the mediator and the DM particle,
the $m_\mathrm{T2}$ distribution of signals may
extend significantly beyond those of backgrounds.

\subsection{ The $2\tau_h+\missET$ Mode}
In the $2\tau_h+\missET$ channel,
SM processes with taus and neutrinos in final states will be
main backgrounds, including diboson ($WW/WZ/ZZ+\mathrm{jets}$),
top pair ($t\bar t+\mathrm{jets}$), and $W+\mathrm{jets}$.
For the $W+\mathrm{jets}$ background, there is a jet mis-tagged as a $\tau_h$.

In order to efficiently suppress backgrounds, we demand the following selection cuts.
\begin{itemize}
\item \textit{Basic cuts:} select the events with $\missET>100~\GeV$ and containing two opposite-sign hadronically decaying taus with $\pT>30~\GeV$ and $|\eta|<2.5$; veto the events containing
any electron ($\pT>10~\GeV$ and $|\eta|<2.47$)
or muon ($\pT>10~\GeV$ and $|\eta|<2.4$).
\item \textit{Jet veto:} veto the events containing
any other central jet ($\pT>30~\GeV$ and $|\eta|<2.5$)
or forward jet ($\pT>40~\GeV$ and $2.5\leq|\eta|<4.5$).
\item \textit{$m_\mathrm{T2}$ cut:}
select the events with $m_\mathrm{T2}>90~\GeV$.
\end{itemize}

\begin{table*}[!htbp]
\centering
\setlength\tabcolsep{0.4em}
\caption{Visible cross sections $\sigma$ (in fb)
and signal significances $\mathcal{S}$
after each cut in the $2\tau_h+\missET$, $\tau_\ell\tau_h+\missET$, and $2\tau_\ell+\missET$ channels at the 14~TeV LHC are presented.
The signal significances are calculated assuming
an integrated luminosity of $3000~\ifb$.}
\label{tab:signif}

\vspace*{.5em}
$2\tau_h+\missET$ channel

\vspace*{.3em}
\begin{tabular}{cccccccccccc}
\hline\hline
 & Diboson & Top pair & $W+\mathrm{jets}$ & \multicolumn{2}{c}{DFDM} & \multicolumn{2}{c}{MFDM} & \multicolumn{2}{c}{CSDM} & \multicolumn{2}{c}{RSDM}\\
 & $\sigma$ & $\sigma$ & $\sigma$ & $\sigma$ & $\mathcal{S}$ & $\sigma$ & $\mathcal{S}$ & $\sigma$ & $\mathcal{S}$ & $\sigma$ & $\mathcal{S}$\\
\hline
Basic cuts & 10.3 & 171 & 182 & 0.0853 & 0.24 & 0.0658 & 0.19 & 0.410 & 1.2 & 0.945 & 2.7 \\
Jet veto & 1.22 & 3.97 & 46.3 & 0.0453 & 0.35 & 0.0349 & 0.27 & 0.220 & 1.7 & 0.499 & 3.8 \\
$m_\mathrm{T2}$ cut & 0.591 & 1.68 & 9.06 & 0.0261 & 0.42 & 0.0214 & 0.35 & 0.167 & 2.7 & 0.336 & 5.4 \\
\hline\hline
\end{tabular}

\vspace*{.5em}
$\tau_\ell\tau_h+\missET$ channel

\vspace*{.3em}
\begin{tabular}{cccccccccccc}
\hline\hline
 & Diboson & Top pair & $W+\mathrm{jets}$ & \multicolumn{2}{c}{DFDM} & \multicolumn{2}{c}{MFDM} & \multicolumn{2}{c}{CSDM} & \multicolumn{2}{c}{RSDM}\\
 & $\sigma$ & $\sigma$ & $\sigma$ & $\sigma$ & $\mathcal{S}$ & $\sigma$ & $\mathcal{S}$ & $\sigma$ & $\mathcal{S}$ & $\sigma$ & $\mathcal{S}$\\
\hline
Basic cuts & 84.3 & 1190 & 1310 & 0.163 & 0.18 & 0.130 & 0.14 & 0.796 & 0.86 & 1.65 & 1.8 \\
Jet veto & 10.6 & 31.0 & 361 & 0.0835 & 0.23 & 0.0674 & 0.18 & 0.424 & 1.2 & 0.811 & 2.2 \\
$m_\mathrm{T2}$ cut & 3.19 & 10.3 & 2.50 & 0.0319 & 0.44 & 0.0293 & 0.40 & 0.263 & 3.6 & 0.372 & 5.0 \\
\hline\hline
\end{tabular}

\vspace*{.5em}
$2\tau_\ell+\missET$ channel

\vspace*{.3em}
\begin{tabular}{ccccccccccc}
\hline\hline
 & Diboson & Top pair & \multicolumn{2}{c}{DFDM} & \multicolumn{2}{c}{MFDM} & \multicolumn{2}{c}{CSDM} & \multicolumn{2}{c}{RSDM}\\
 & $\sigma$ & $\sigma$ & $\sigma$ & $\mathcal{S}$ & $\sigma$ & $\mathcal{S}$ & $\sigma$ & $\mathcal{S}$ & $\sigma$ & $\mathcal{S}$\\
\hline
Basic cuts & 918 & 5660 & 0.115 & 0.078 & 0.0884 & 0.060 & 0.526 & 0.36 & 1.29 & 0.87 \\
Jet veto & 354 & 204 & 0.0629 & 0.15 & 0.0483 & 0.11 & 0.291 & 0.67 & 0.698 & 1.6 \\
$Z$ veto & 281 & 190 & 0.0598 & 0.15 & 0.0462 & 0.12 & 0.275 & 0.69 & 0.655 & 1.7 \\
$m_\mathrm{T2}$ cut & 0.500 & 0.388 & 0.00593 & 0.34 & 0.00649 & 0.38 & 0.0681 & 3.8 & 0.0549 & 3.1 \\
\hline\hline
\end{tabular}
\end{table*}

Tab.~\ref{tab:signif} lists the visible cross sections of backgrounds
and signals after each cut. The visible cross section is defined as
the production cross section times cut acceptance and efficiency.
The benchmark points of signals are chosen as
$m_\phi=225~(250)~\GeV$ for the DFDM (MFDM) and
$m_\psi=300~(200)~\GeV$ for the CSDM (RSDM).
We define the signal significance by $\mathcal{S}=S/\sqrt{B+S}$,
where $S$ ($B$) is the number of signal (total background) events.
In order to exhibit the cut efficiency,
we also list the values of $\mathcal{S}$ for the benchmark points
with an integrated luminosity of $3000~\ifb$.

After imposing the basic cuts,
the top pair and $W+\mathrm{jets}$ backgrounds are dominant.
The jet veto is particularly powerful for suppressing the top pair background,
where two $b$ quarks from top decays usually give rise to two jets.
The $m_\mathrm{T2}$ distributions of the two tagged taus
for backgrounds and signals are demonstrated in Fig.~\ref{fig:mt2:2tau}.
The distribution for the $W+\mathrm{jets}$ background
drops quickly as $m_\mathrm{T2}$ increases,
and the $m_\mathrm{T2}$ cut can kill 80\% of this background.
Due to fake taus and the imperfect energy measurement
caused by $\tau$-decay neutrinos,
however, the $m_\mathrm{T2}$ distributions for the signals
seem not quite separate from those for the diboson and top pair backgrounds.
A higher threshold of the $m_\mathrm{T2}$ cut is not applied,
otherwise too much signal events will be lost.

\subsection{ The $\tau_\ell \tau_h+\missET$ Mode}

In the $\tau_\ell\tau_h+\missET$ channel,
main backgrounds are similar to those in the $2\tau_h+\missET$ channel.
We use the following selection cuts.
\begin{itemize}
\item \textit{Basic cuts:} select the events with $\missET>100~\GeV$ and containing
one hadronically decaying tau ($\pT>30~\GeV$ and $|\eta|<2.5$)
and one light lepton (one electron with $\pT>20~\GeV$ and $|\eta|<2.47$,
or one muon with $\pT>20~\GeV$ and $|\eta|<2.4$).
We further require that the signs of the reconstructed tau and the light lepton are opposite.
\item \textit{Jet veto:} veto the events containing
any other central jet ($\pT>30~\GeV$ and $|\eta|<2.5$)
or forward jet ($\pT>40~\GeV$ and $2.5\leq|\eta|<4.5$).
\item \textit{$m_\mathrm{T2}$ cut:}
select the events with $m_\mathrm{T2}>90~\GeV$.
\end{itemize}

The visible cross sections of backgrounds and signals
as well as the signal significances for the benchmark points
in this channel are also presented in Tab.~\ref{tab:signif}.
After the basic cuts and jet veto, the $W+\mathrm{jets}$ background
is dominant. We use the light lepton and the reconstructed $\tau_h$
to construct the $m_\mathrm{T2}$ variable, whose distributions
are shown in Fig.~\ref{fig:mt2:1lep1tau}.
The $m_\mathrm{T2}$ distribution for $W+\mathrm{jets}$
essentially ends at around the $W$ mass.
It is observed that the $m_\mathrm{T2}$ variable
works as efficiently as the ordinary $m_\mathrm{T}$ variable,
which is constructed by $\missET$ and the lepton $\mathbf{p}_\mathrm{T}$
and also give an end point for this background.
After the $m_\mathrm{T2}$ cut, the $W+\mathrm{jets}$ background is suppressed by two orders of magnitude and is less dominant compared with the top pair background.

\subsection{The $2 \tau_\ell +\missET$ Mode}

In the $2\tau_\ell+\missET$ channel,
main backgrounds are di-boson and top pair production.
The following selection cuts are adopted.
\begin{itemize}
\item \textit{Basic cuts:}
select the events with $\missET>60~\GeV$ and containing two opposite-sign light leptons (electrons with $\pT>20~\GeV$ and $|\eta|<2.47$, muons with $\pT>20~\GeV$ and $|\eta|<2.4$);
veto the events containing any hadronically decaying tau
with $\pT>30~\GeV$ and $|\eta|<2.5$.
\item \textit{Jet veto:} veto the events containing
any central jet ($\pT>30~\GeV$ and $|\eta|<2.5$),
or forward jet ($\pT>40~\GeV$ and $2.5\leq|\eta|<4.5$).
\item \textit{$Z$ veto:} if the two light leptons have the same flavor,
their invariant mass must satisfy $|m_{\ell\ell}-m_Z|>10~\GeV$.
\item \textit{$m_\mathrm{T2}$ cut:}
select the events with $m_\mathrm{T2}>100~\GeV$.
\end{itemize}

\begin{figure*}[!htbp]
\centering
\subfigure[~$2\tau_h+\missET$ channel.\label{fig:mt2:2tau}]
{\includegraphics[width=.42\textwidth]{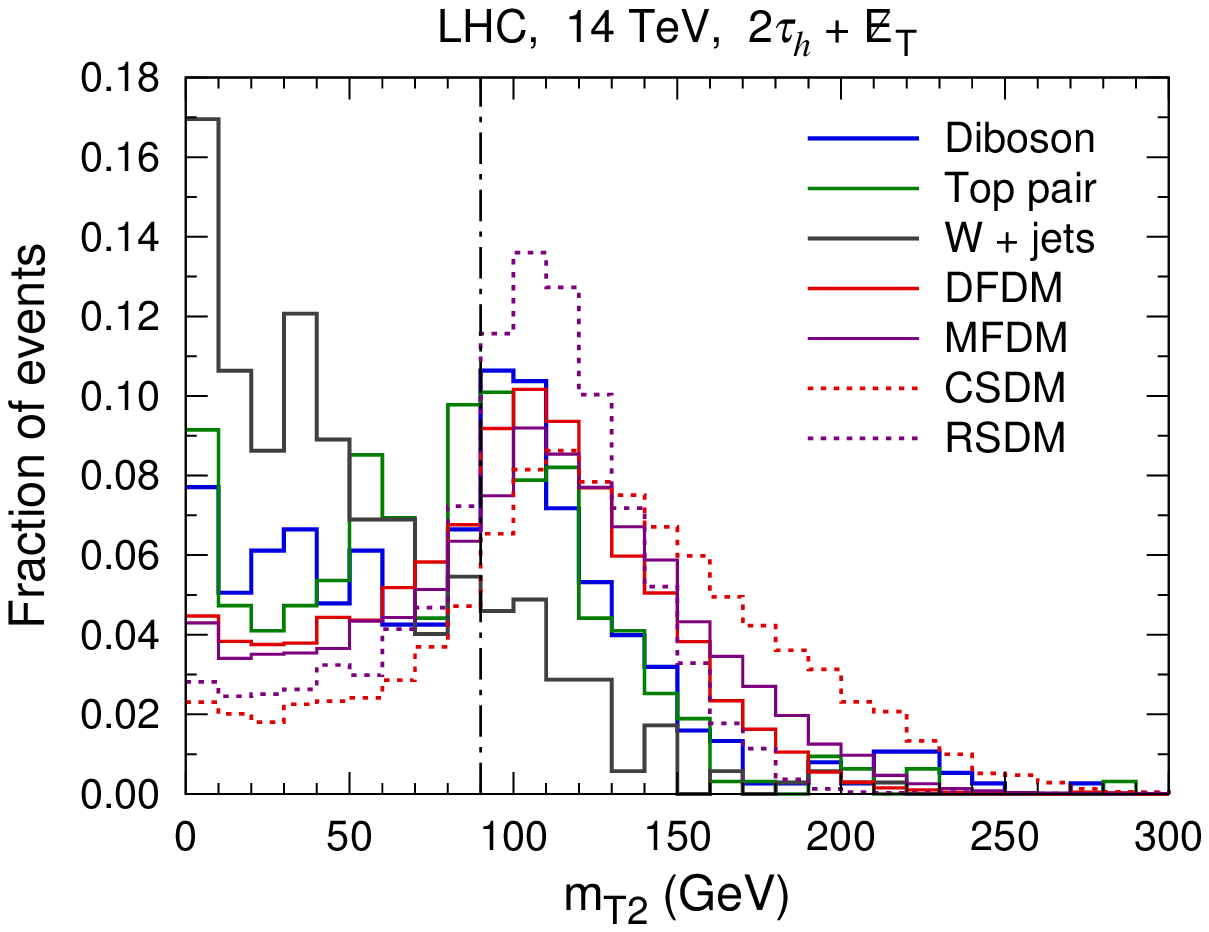}}
\subfigure[~$\tau_\ell\tau_h+\missET$ channel.\label{fig:mt2:1lep1tau}]
{\includegraphics[width=.42\textwidth]{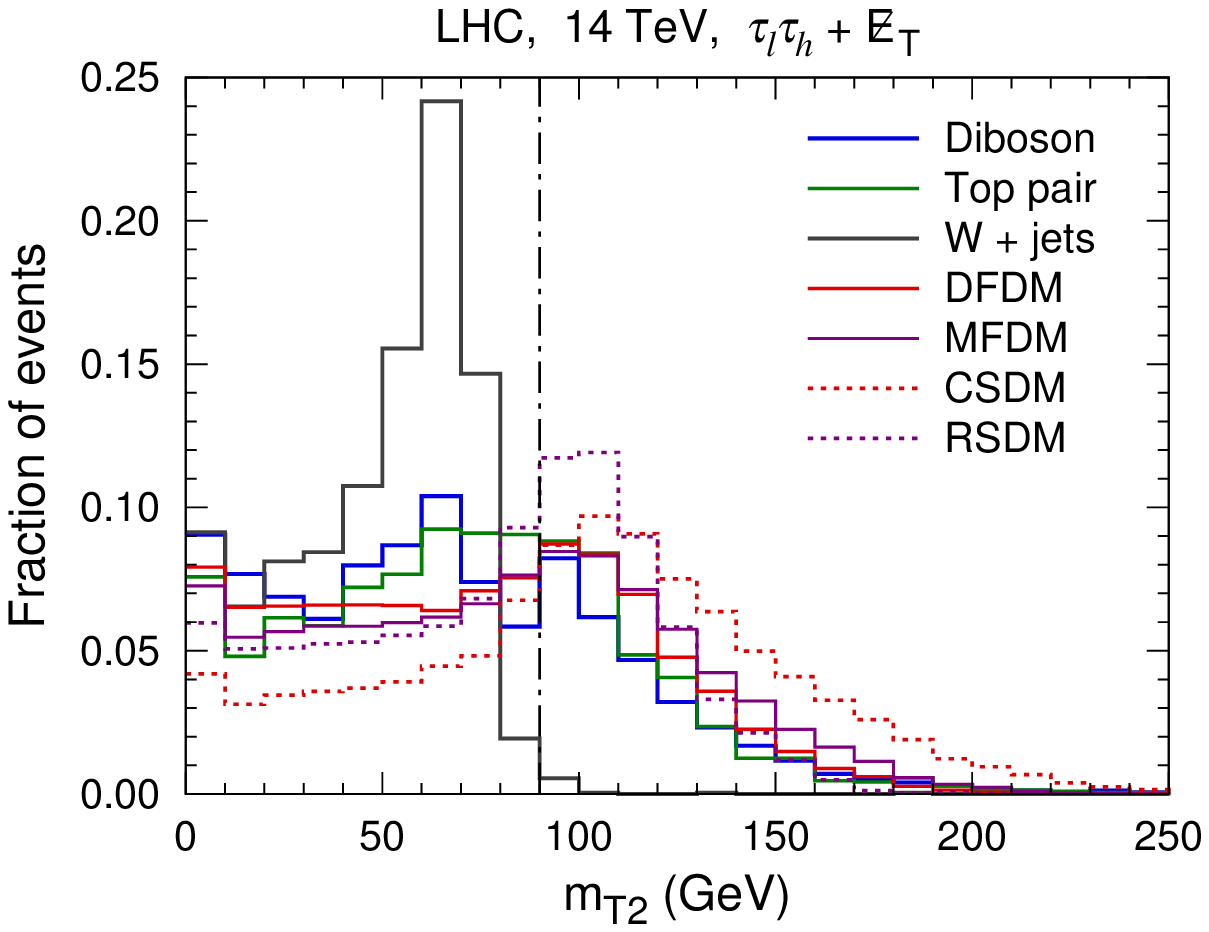}}
\subfigure[~$2\tau_\ell+\missET$ channel.\label{fig:mt2:2lep}]
{\includegraphics[width=.42\textwidth]{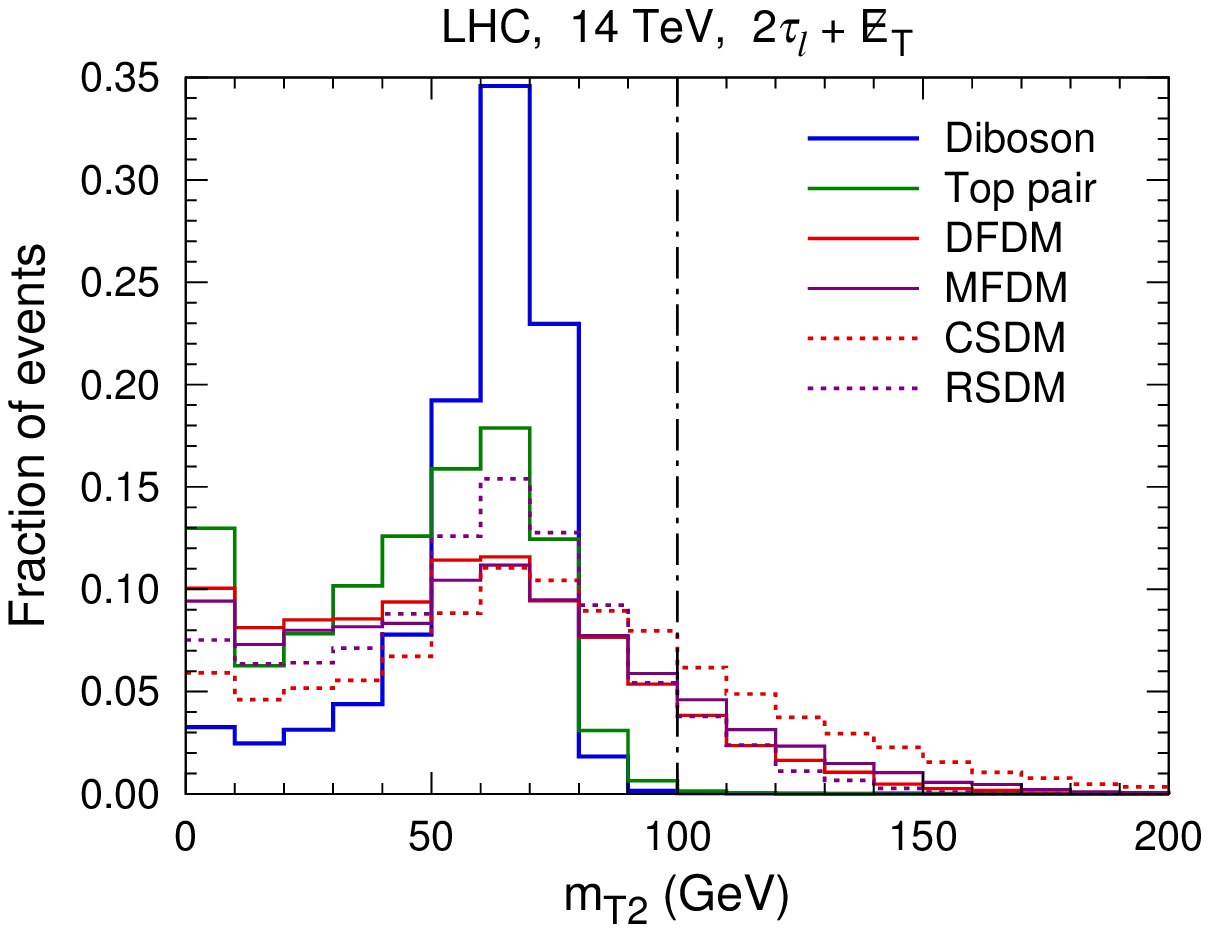}}
\caption{Normalized $m_\mathrm{T2}$ distributions for backgrounds and signals
in the $2\tau_h+\missET$ (a), $\tau_\ell\tau_h+\missET$ (b),
and $2\tau_\ell+\missET$ (c) channels at the 14~TeV LHC are demonstrated.
The dot-dashed vertical lines denote the locations of
the $m_\mathrm{T2}$ cuts.}
\label{fig:mt2}
\end{figure*}

Tab.~\ref{tab:signif} also demonstrates the visible cross sections
and signal significances in this channel.
Although the branching fraction of the $2\tau_\ell+\missET$ channel is small,
a little more signal events remain after the basic cuts
compared with the $2\tau_h+\missET$ channel.
The top pair background is again significantly suppressed by the jet veto.
In the invariant mass distribution of two opposite-sign same-flavor
light leptons, the diboson background has a peak at the $Z$ mass pole,
which will be removed by the $Z$ veto.
Fig.~\ref{fig:mt2:2lep} shows the $m_\mathrm{T2}$ distributions
of the two light leptons. The $m_\mathrm{T2}$ distributions for the backgrounds
exhibit clear endpoints at around the $W$ mass,
while the distributions for the signals extend to higher values.
As a result, the $m_\mathrm{T2}$ cut is extremely powerful to suppress the backgrounds.

\begin{figure*}[!htbp]
\centering
\subfigure[~$2\tau_h+\missET$ channel.]
{\includegraphics[width=.42\textwidth]{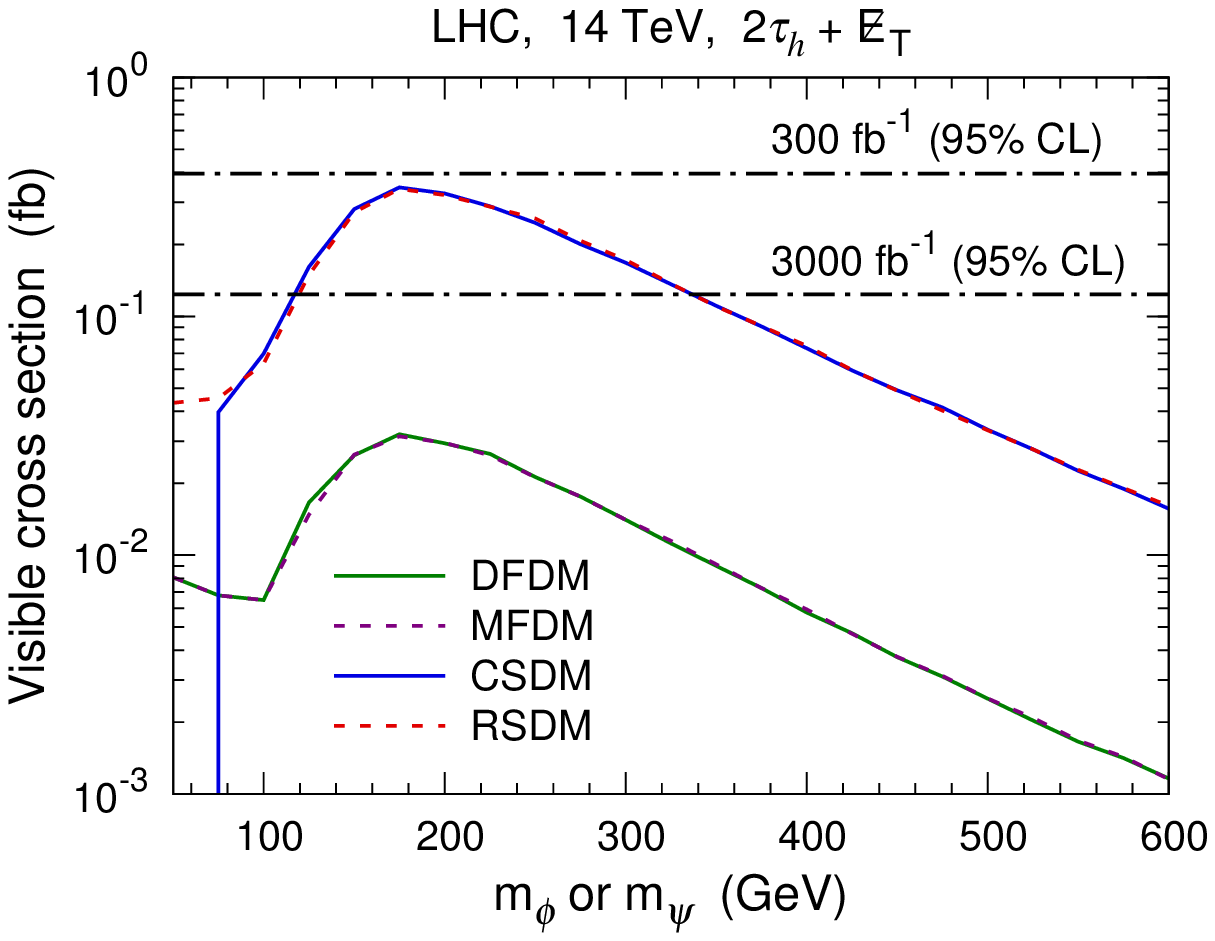}}
\subfigure[~$\tau_\ell\tau_h+\missET$ channel.]
{\includegraphics[width=.42\textwidth]{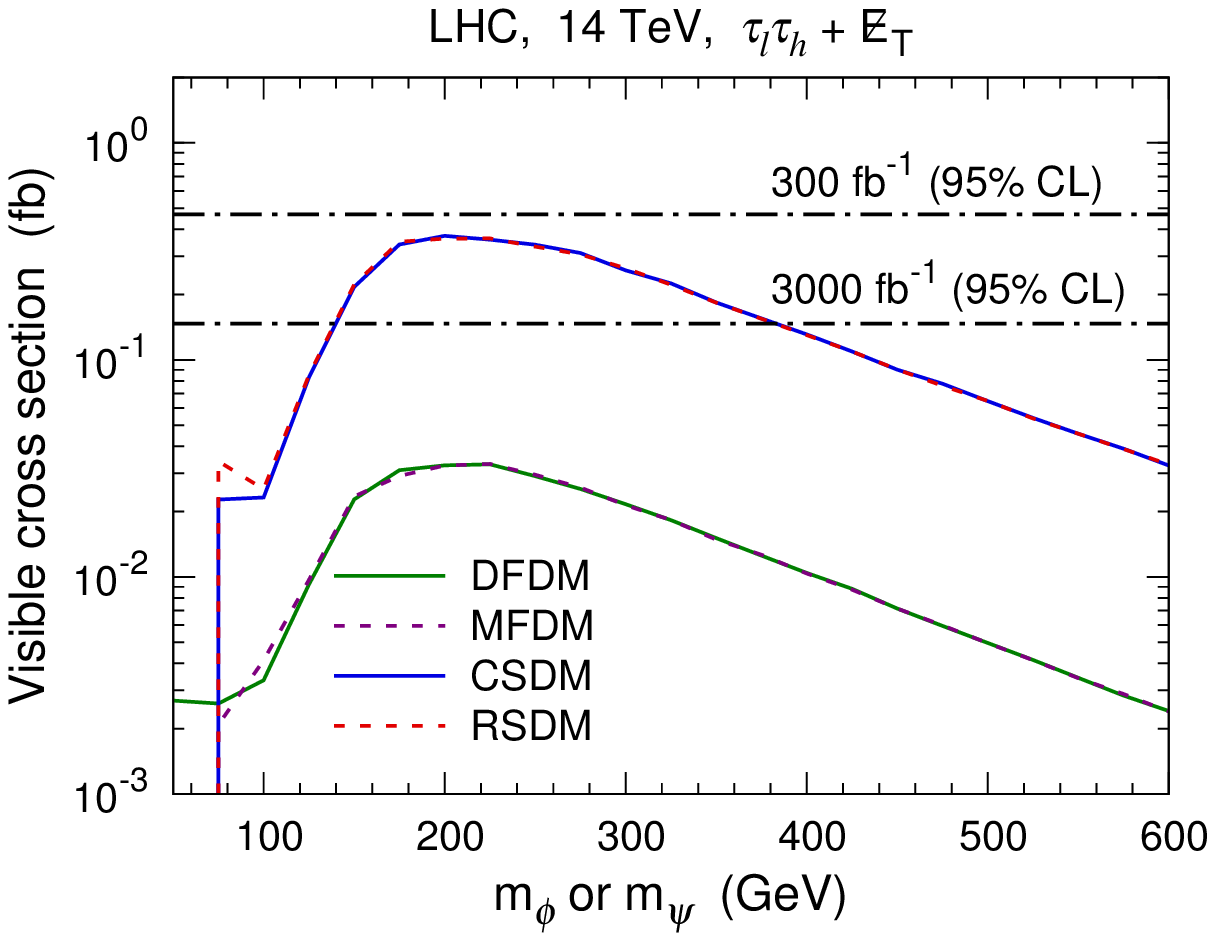}}
\subfigure[~$2\tau_\ell+\missET$ channel.]
{\includegraphics[width=.42\textwidth]{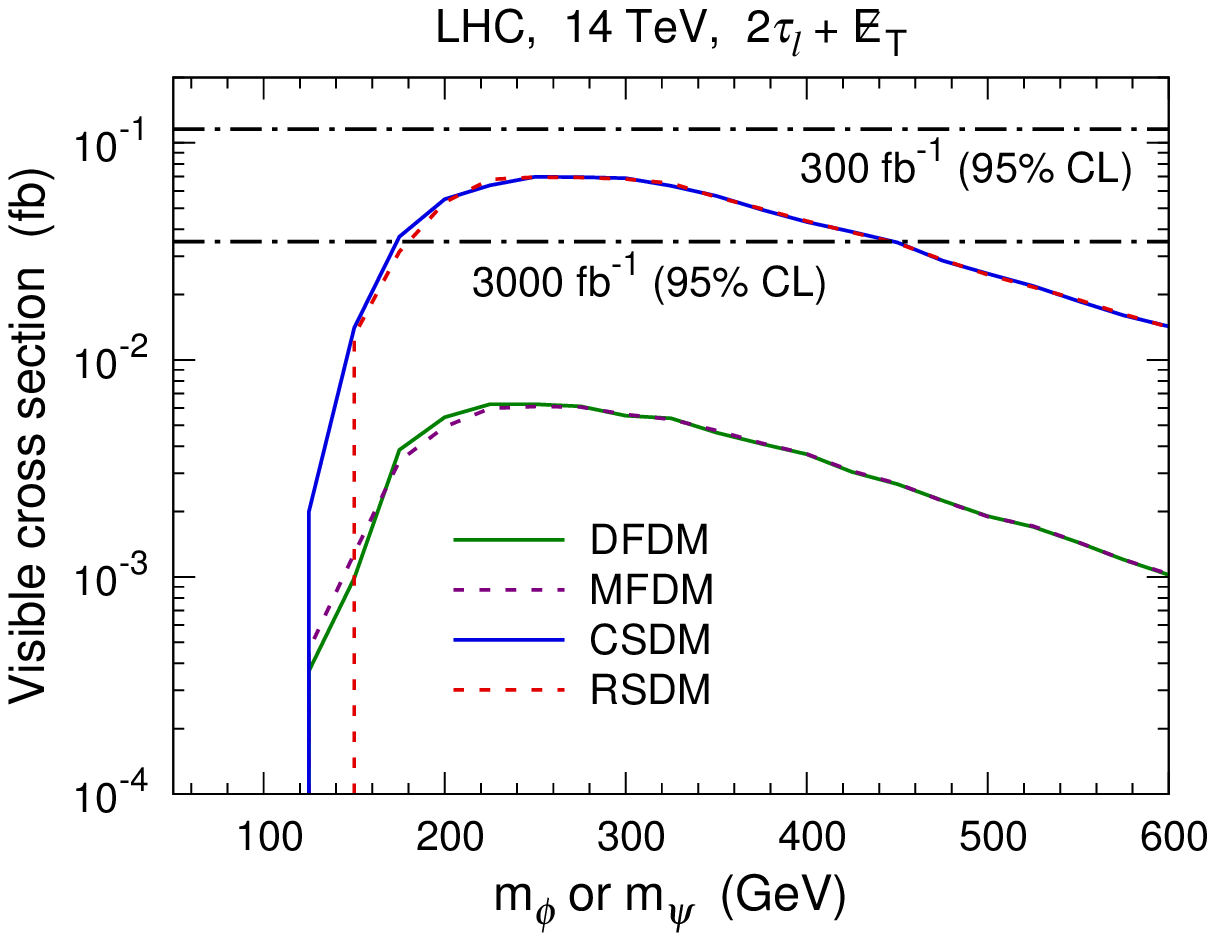}}
\caption{Visible cross sections after all the cuts
in the $2\tau_h+\missET$ (a), $\tau_\ell\tau_h+\missET$ (b),
and $2\tau_\ell+\missET$ (c) channels
for the tau portal DM models at the 14~TeV LHC are shown.
The dot-dashed horizontal lines denote the expected 95\% C.L. exclusion limits
assuming integrated luminosities of $300~\ifb$ and $3000~\ifb$, respectively.}
\label{fig:vis_Xsec}
\end{figure*}

The visible cross sections after applying all the cuts
as functions of $m_\phi$ or $m_\psi$
are shown in Fig.~\ref{fig:vis_Xsec} for the tau portal DM models.
Due to the same production cross sections and event distributions,
the results for the MFDM and RSDM are basically identical to
those for the DFDM and CSDM, respectively.
The slight discrepancies at low masses are caused by
the fluctuations of simulation samples.
We also show the expected 95\% C.L. exclusion limits
at the 14~TeV LHC with integrated luminosities of $300~\ifb$ and $3000~\ifb$.
It is observed that a date set of $300~\ifb$ seems not sufficient to explore these models. In contrast, the high luminosity LHC (HL-LHC) with an integrated luminosity of $3000~\ifb$ is capable to explore the CSDM and RSDM models.
The sensitivities of the three searching channels are similar but not identical.
The $2\tau_\ell+\missET$ channel can cover a wider mediator mass range,
while the $2\tau_h+\missET$ channel can achieve higher significances
at its sensitive mediator masses. The $\tau_\ell\tau_h+\missET$ channel
seems to give an average sensitivity between the other two channels.
On the other hand, the DFDM and MFDM models,
due to the much lower production rates,
can hardly be constrained even at the HL-LHC.

\begin{figure*}[!htbp]
\centering
\subfigure[~Dirac fermionic dark matter.
\label{fig:constr:DFDM}]
{\includegraphics[width=.42\textwidth]{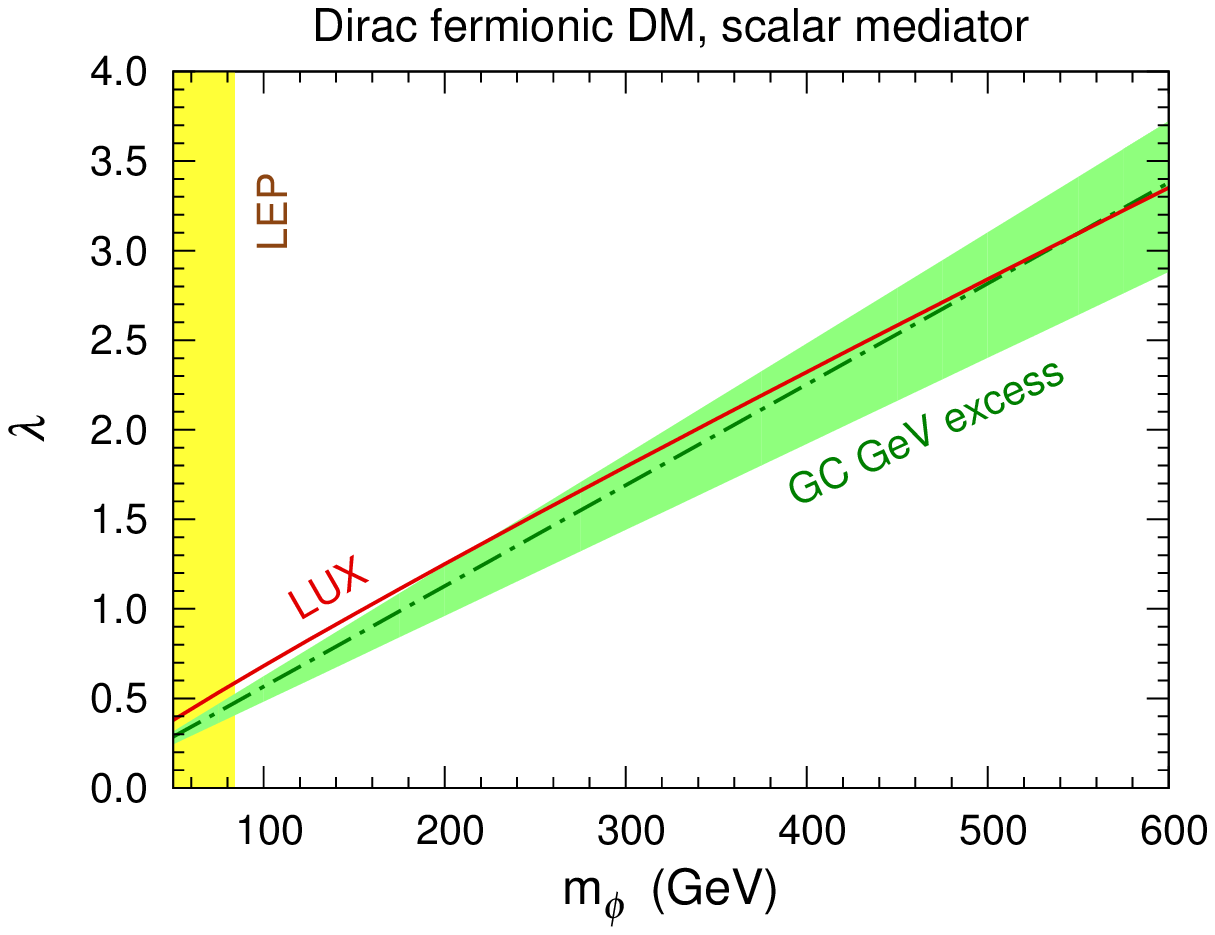}}
\subfigure[~Majorana fermionic dark matter.
\label{fig:constr:MFDM}]
{\includegraphics[width=.42\textwidth]{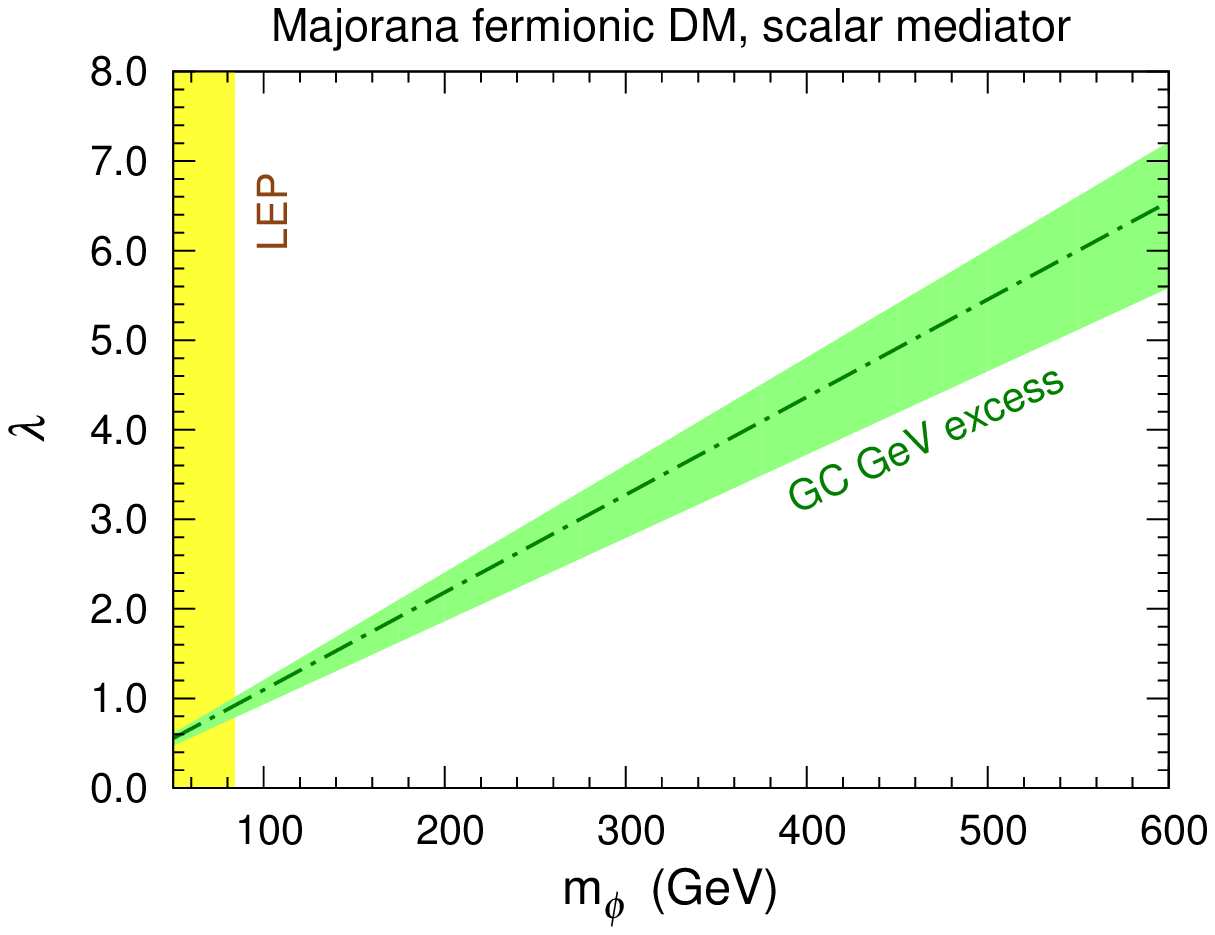}}
\subfigure[~Complex scalar dark matter.
\label{fig:constr:CSDM}]
{\includegraphics[width=.42\textwidth]{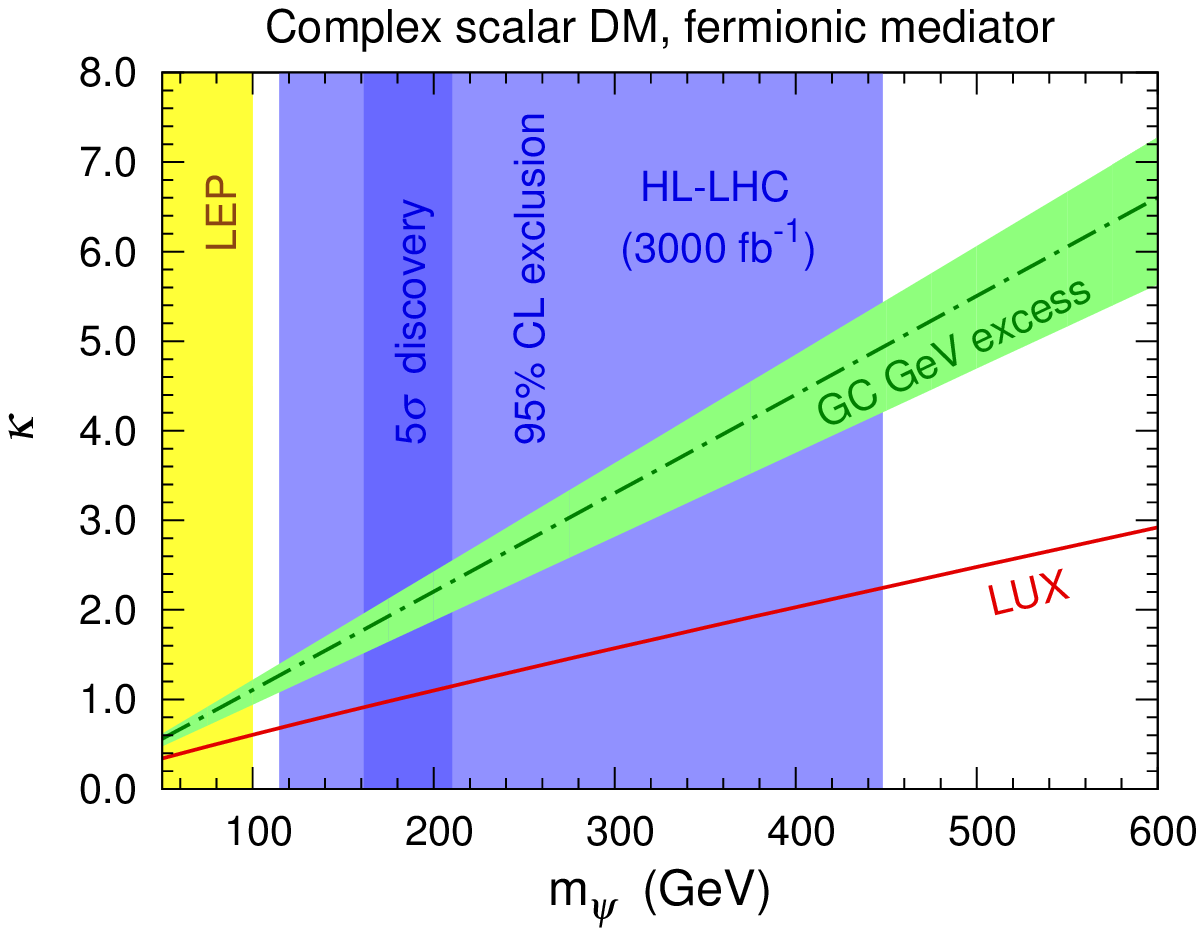}}
\subfigure[~Real scalar dark matter.
\label{fig:constr:RSDM}]
{\includegraphics[width=.42\textwidth]{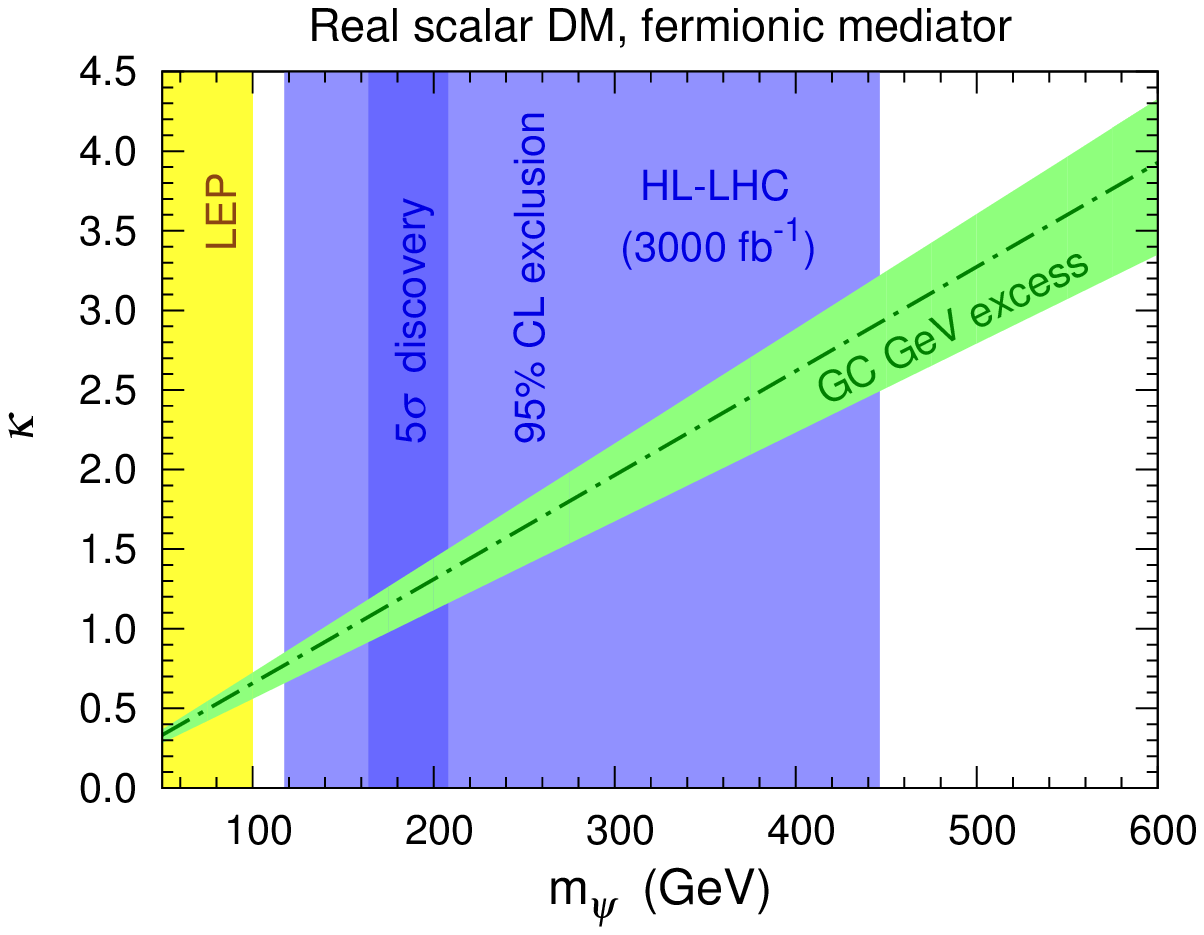}}
\caption{Parameter spaces of the DFDM (a), MFDM (b), CSDM (c), and RSDM (d) models for $m_\chi=9.43~\GeV$
are shown. The green bands correspond to the annihilation cross section
consistent with the GC gamma-ray excess within $1\sigma$
uncertainty~\cite{Abazajian:2014fta},
while the dot-dashed lines denote the central value.
The red solid lines represent the 90\% C.L. exclusion limit given by
the LUX direct detection experiment~\cite{Akerib:2013tjd}.
The yellow exclusion regions are inferred from the DELPHI searching results
at the LEP~\cite{Abdallah:2003xe}.
The expected 95\% C.L. exclusion ($5\sigma$ discovery) regions estimated by
combining all the three channels
at the HL-LHC with an integrated luminosity of $3000~\ifb$
are demonstrated by the light (dark) blue color.}
\label{fig:constr}
\end{figure*}

In Figs.~\ref{fig:constr:CSDM} and \ref{fig:constr:RSDM},
we demonstrate the sensitive parameter regions for the CSDM and RSDM models
by combining all the three searching
channels at the HL-LHC with an integrated luminosity of $3000~\ifb$.
The expected 95\% C.L. exclusion ($5\sigma$ discovery) regions can cover
the range $120~\GeV\lesssim m_\psi\lesssim 450~\GeV$
($160~\GeV\lesssim m_\psi\lesssim 210~\GeV$).

\section{Conclusions and discussions}
\label{sec:concl}

In this work, we consider a class of tau portal dark matter simplified models
to explain the Galactic Center gamma-ray excess
and study current constraints from the direct detection and
future prospects of LHC searches.
The LUX experiment has ruled out the region favored
by the gamma-ray excess in the CSDM model,
and has also constrained the DFDM model for $m_\phi\gtrsim 200~\GeV$.
In order to further test these models at the LHC, we propose three searching channels for $\tau^+ \tau^- +\missET$ signatures,
which are challenging due to the small rates of electroweak productions.
The $m_\mathrm{T2}$ variable is utilized to suppress backgrounds.
Based on the cut-based method, we expect that the HL-LHC searches
can explore the RSDM model up to $m_\psi\sim 450~\GeV$.
Except for the limit $m_\phi\gtrsim 84~\GeV$ set by the LEP searches,
the MFDM model is unlikely to be tested by
direct detection experiments and LHC searches,
making it a task for future high energy $e^+e^-$ colliders,
such as TLEP, CEPC, and ILC.

The sensitivity at the 14~TeV LHC we estimate is based on
a simplified simulation, which captures the main features
but ignores some subtle issues.
For instance, some minor backgrounds are not considered,
and the identification efficiency and faking rate of
the $\tau$-tagging technique are assumed to be flat.
Moreover, systematic uncertainties are not included
when estimating the significance.
In spite of these simplifications, the study should be considered
as a first order approximation and can be a useful guide for the real experimental searches.
In addition, it is worth mentioning that a better performance of
the $\tau$-tagging technique, which may be achieved at the 14~TeV LHC,
could undoubtedly improve the sensitivities of the $2\tau_h+\missET$ and $\tau_\ell\tau_h+\missET$ channels.

In the future, high energy $e^+e^-$ colliders, such as CEPC, TLEP and ILC, are powerful to search for the charged mediators in these tau portal DM models. Backgrounds at $e^+e^-$ colliders can be well handled. Once the mediators are pair-produced, the signature of $\tau^+ \tau^- +\missE$ would be reconstructed quite precisely. Therefore, the sensitivity of $e^+e^-$ colliders to the mediators dominantly depends on the collision energy. For instance, it is possible to test or exclude the masses of these mediators up to $\sim500$~GeV at the ILC with $\sqrt{s}=1$~TeV.

\begin{acknowledgments}
This work is supported by the Natural Science Foundation of China under Grants No.~11105157, No.~11175251, and No.~11135009, and the 973 Program under Grant No.~2013CB837000.
\end{acknowledgments}


\end{document}